# $C_n^2$ and wind profiler method to quantify the frozen flow decay using wide-field laser guide stars adaptive optics

Andrés Guesalaga[1*], Benoit Neichel[2,3], Angela Cortes[1], Clémentine Béchet[1], Dani Guzmán[1]

[1] Pontificia Universidad Católica de Chile, 4860 Vicuña Mackenna, Casilla 7820436, Santiago, Chile.
[2] Gemini Observatory Southern Operations Center, Colina el Pino s/n, Casilla 603, La Serena, Chile.
[3] Aix Marseille Université, CNRS, LAM (Laboratoire d'Astrophysique de Marseille) UMR 7326, 13388 Marseille, France.

* E-mail: aguesala@ing.puc.cl

## ABSTRACT

We use spatio-temporal cross-correlations of slopes from five Shack-Hartmann wavefront sensors to analyse the temporal evolution of the atmospheric turbulence layers at different altitudes. The focus is on the verification of the frozen flow assumption. The data is coming from the Gemini South Multi-Conjugate Adaptive Optics System (GeMS).

First, the $C_n^2$ and wind profiling technique is presented. This method provides useful information for the AO system operation such as the number of existing turbulence layers, their associated velocities, altitudes and strengths and also a mechanism to estimate the dome seeing contribution to the total turbulence.

Next, by identifying the turbulence layers we show that it is possible to estimate the rate of decay in time of the correlation among turbulence measurements. We reduce on-sky data obtained during 2011, 2012 and 2013 campaigns and the first results suggest that the rate of temporal de-correlation can be expressed in terms of a single parameter that is independent of the layer altitude and turbulence strength.

Finally, we show that the decay rate of the frozen-flow contribution increases linearly with the layer speed. The observed evolution of the decay rate confirms the potential interest of the predictive control for wide-field AO systems.

## 1. INTRODUCTION

The goal of adaptive optics (AO) in astronomy is to correct the phase aberrations that are caused by the variation of the refractive index of the atmosphere. The frozen flow assumption or Taylor's hypothesis (Taylor, 1938) states that these variations stay spatially stable for a time scale comparable to the time it takes the turbulence to transverse the field of view (FoV) of the telescope. Although this assumption is generally used in predictive control and in AO modelling and simulation, only a few experimental studies to validate this hypothesis can be found in the literature (Gendron and Lena 1996; Schöck and Spillar 2000; Britton, 2004; Guyon, 2005; Poyneer *et al*, 2009). Results about the time-scales in which the wavefront can be predicted under this assumption are not unanimous, but values of tens of milliseconds (Schöck and Spillar, 2000) are generally accepted. Regarding the stability of the velocity vector for each layer, the time scales are much larger, ranging from seconds to several minutes and even hours (Poyneer *et al*, 2009).

In astronomical AO systems, there is a time delay between the start of the wavefront sensor (WFS) measurement and the response of the deformable mirror (DM) to the actuator computed from these measurements. This delay reduces the performance of the AO loop in what is known as the temporal error caused by the evolution of phase during this delay. Predictive control (see for instance Gavel and Wiberg, 2002; Le Roux *et al*, 2004; Hinnen *et al*, 2007 and Poyneer *et al*, 2007) can improve the performance of AO systems substantially if the frozen flow assumption holds and it actually contributes to the phase aberration corrections at the different layers.

In this paper we analyse the characteristics of the frozen-flow, in order to determine the dynamics of the temporal de-correlation ("boiling") experienced by the turbulence layers (Saint-Jacques and Baldwin, 2000; Jolissaint, 2006; and Berdja and Borgnino, 2007). Our results are based on observations taken from the Gemini South multi-conjugate AO system (GeMS), which is described in section 2. In section 3, we present the method to post-process GeMS telemetry data to determine the turbulence $C_n^2$ and wind profiles and also use them to study and verify the frozen flow assumption. In this section, the choice of the method for the profile estimation is also discussed, i.e. model fitting of specific turbulence laws (e.g. Kolmogorov or von Kármán) or cross-correlations of slopes deconvolved by the measured autocorrelations. We present our results on frozen flow analysis in section 4. We find that the rate of decay in the temporal cross-correlation of the layers follow a very simple rule given by the distance travelled across the telescope's field of view. Section 5 gives the conclusions.

## 2. TURBULENCE PROFILING & GeMS DATA

### 2.1 Turbulence profiling methods

The turbulence profiling can be obtained from Shack-Hartmann data by SLODAR-like methods (Wilson, 2002). These methods are based on optical triangulation for the measurement of the atmospheric optical turbulence profile, using the spatial or spatio-temporal cross-correlations of the slopes measured by the WFSs, each pointing at different guide stars. The turbulence strength can be estimated in as many altitude bins as subapertures across the WFSs exist and the SLODAR method can be adapted to use Laser Guide Stars (LGS) as previously reported (Fusco and Costille 2010; Gilles and Ellerbroek 2010; Osborn *et al* 2012; Cortés *et al*, 2012). In such cases, the turbulence profiling with LGS is performed to non-equally spaced bin altitudes due to the cone effect. As an extension to the SLODAR method, the wind profiling consists in performing time-delayed cross-correlations between all possible combinations of wavefront measurements from the available WFSs data (Wang, Schöck and Chanan, 2008). This profiling method provides information not only on the turbulence distribution in altitude but also on the wind velocity for each layer and their dynamic evolution. The method has also been modified to include multiple LGS-WFSs and the handling of the cone and fratricide effects (Cortes *et al*, 2012).

For illustrative purposes, Fig. 1 shows a simple configuration based on only two LGS to construct a turbulence profile ($C_n^2(h_m)$). Here, $h_m$ ($m$=0,...,$N$-1) is the altitude of each bin. The finite distance to the stars means that the light from the guide stars forms a cone. This cone effect reduces the area illuminated by the guide star at higher altitudes.

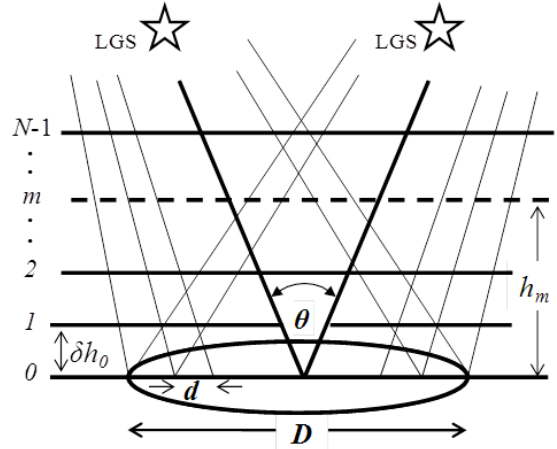

**Figure 1.** Two laser star configuration for a WFS with *N* x *N* subapertures. *D* is the telescope pupil diameter and $\theta$ is the angular separation between the stars.

## 2.2 Description of the data

For altitude turbulence profiling, multiple WFS data are used. In our case, the data come from GeMS, a facility instrument that delivers a uniform, diffraction limited image quality at near-infrared wavelengths over an extended FoV of more than 1 arcmin across (Rigaut *et al*. 2013). This corrected beam then feeds the science instruments. GeMS consist of 3 main subsystems: i) the laser source that provides enough power to generate 5 laser beacons in the sodium layer; ii) the beam transfer optics that takes the laser beam from its source, splits it into five beacons and transfers it to the laser launch telescope (behind the secondary mirror) and iii) the AO bench where the wavefront sensors and correctors are located.

The beam from the telescope is taken to GeMS' AO bench, where it is collimated onto two DMs conjugated at different altitudes (0 and 9 km respectively) and a tip-tilt mirror (TTM). A science beam splitter transmits the infrared light to the science path, and the 589 nm wavelength from the five laser beacons are reflected by the LGS beam splitter and sent to the LGS WFSs. Each WFS is a Shack-Hartmann of 16x16 with 204 valid subapertures, resulting to 2040 values of slopes (axis X and Y) sampling at a maximum frequency of 800 Hz. The pixel size of the WFS is about 1.38'' and the measured read out noise is on the order of 3.5 *e*. Each subaperture on the CCD uses 2x2 pixels (quadcell). For a detailed description of this system the reader is referred to Neichel et al (2012).

The 2040 slopes from GeMS WFSs are initially available for the profile estimation. However, in order to eliminate noisy or low reliability subaperture slopes, these resulting slopes are subject to a masking procedure (Cortes *et al.* 2012). Figure 2 shows the photon intensity received at the five WFSs where the brighter dots correspond to pixels contaminated by the Rayleigh scattering of adjacent lasers during propagation (Neichel et al. 2011). To mitigate the impact of the fratricide effect and of the partial illumination of some subapertures in the outer pupil rings, some slopes are systematically eliminated via masks applied to the WFS, leading to a final set of 1280 slopes instead of the initial 2040 (see Fig. 3).

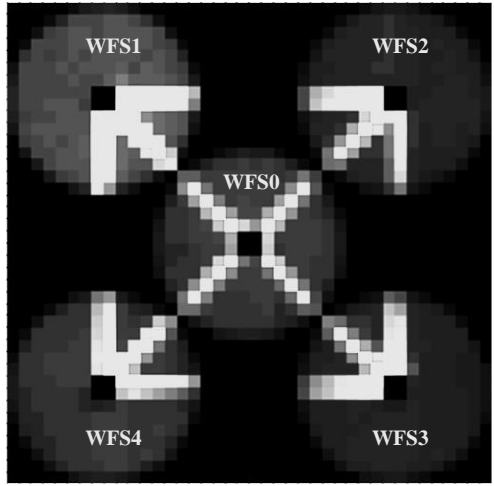

**Figure 2.** The fratricide effect: photon intensity at the WFSs

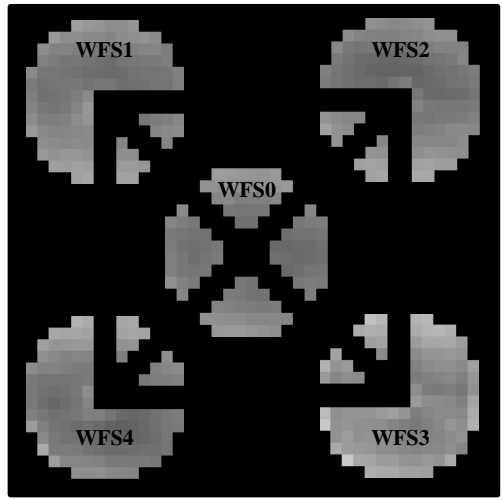

**Figure 3.** Variance of POL-slopes in the X direction. A fairly uniform variance is observed among the five WFSs. Black areas are masks applied to the non-valid subapertures contaminated with the fratricide effect or noise caused by partial illumination of the outer ring (data from April 15th, 2011).

Since GeMS normally operates in closed-loop and the profiling technique works with open-loop data, it is necessary to estimate the original slopes of the incoming wavefront. This is done through the pseudo-open-loop (POL) reconstruction process and consists of adding the slopes of the residuals $S^{res}$ to the DMs voltages $V^{act}$ projected onto the slope domain by means of the interaction matrix (*iMat*) that corresponds to the static response of an AO system. This can be represented by the following equation:

$$S_i^{pol} = S_i^{res} + iMat \cdot V_{i-1}^{act} \quad , \qquad (1)$$

where *i* is the discrete time. A 1-frame delay exists between the voltages and the slopes due to the exposure and readout time of the CCD. The data collected during runs are stored in circular buffers. For the profiler we require the slopes of the AO loop residuals ($S^{res}$) and the corresponding actuators voltages ($V^{act}$). Data are saved at the loop frame rate, which can range from 100Hz to 800Hz. An individual circular buffer covers a period of 30 seconds to 4 minutes. Circular buffers are saved regularly during operations.

The time-averaged centroids and piston voltages are subtracted to remove biases and bad actuators. Also, any common global motion in each WFS (tip-tilt) is subtracted from the corresponding slopes, so as to remove wind-shake and guiding errors. Focus removal is unnecessary, as GeMS compensates for sodium altitude fluctuations using a slow focus sensor (Neichel *et al*, 2012).

Finally, for a LGS asterism with a "X" shape as in GeMS (Neichel *et al*, 2010), up to 10 pair combinations exist for the star asterism shown in Fig. 4. These pairings provide two different altitude resolutions; a low resolution and long estimation range and a high resolution with a shorter profile range (Cortés *et al*, 2012).

By combining the two types of resolutions, an extended range with reliable profile estimations can be achieved, reaching distances of over 20 Km.

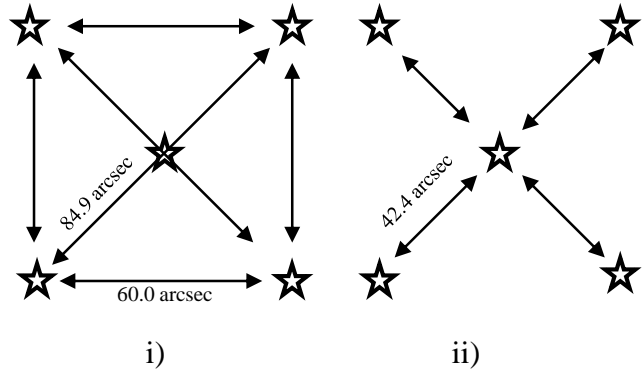

**Figure 4.** Three angular separation of LGSs giving two altitude resolutions. Combinations i) correspond to high resolution profiles whereas ii) gives low resolution and higher altitude information (Cortés *et al*, 2012).

## 3. ATMOSPHERE TURBULENCE PROFILING USING TIME-DELAYED CROSS-CORRELATION

### 3.1. Methods for turbulence profiling

In the literature, two different approaches have been used to analyse the data from SLODAR optical triangulation. The first approach deduces the $C_n^2$ and wind profiles from the spatio-temporal cross-correlations deconvolved by the auto-correlation of the data (Wilson, 2002; Wang, Schöck and Chanan, 2008). The second one does not apply such deconvolution, but rather models the spatial covariances of the slopes using a specific turbulence statistics (e.g. Kolmogorov or von Kármán): the turbulent layers heights and strengths are recovered by fitting theoretical impulse response functions to the cross-covariance of the slopes measurements (Butterley, Wilson, and Sarazin, 2006).

The model fitting approach has been claimed by Butterley *et al* (2006) to be more accurate than the deconvolution. However, in many cases encountered in GeMS data, the profiles obtained by the model fitting differ significantly from the deconvolution approach (Guesalaga *et al*, 2013).

For data collected with GeMS over a two-year period, strong dome seeing conditions were detected in a significant number of cases (~ 30% of the recorded conditions). It has been observed that the statistics of such turbulence contribution narrows the auto-covariance impulse response causing a clear departure from Kolmogorov or von Kármán theoretical shape.

This discrepancy suggests that in the model fitting approach, a different turbulence statistics should be used for fitting the ground layer bin. Nevertheless, the physics and statistical law of the turbulence in the telescope dome is, as for the turbulence of the surface layer, still a matter of research (Lombardi *et al.* 2010). Since no clear statistical model exist so far, trying to fit the measured covariance submaps to these theoretical statistical models can lead to erroneous profile estimations (Guesalaga *et al*, 2013).

On the other hand, the use of the deconvolved cross-correlation implicitly accounts for the contribution of dome turbulence as the deconvolution function is constructed from the actual measurements. When this dome contribution is large, a narrower convolution function is obtained, leading to more realistic estimates of the lower layers. However, this autocorrelation function also includes the turbulence from the upper layers, which should behave closer to Kolmogorov statistics. This loss of precision of the deconvolution approach for upper layers was already mentioned by Butterley *et al* (2006).

In conclusion, we privilege the use of the deconvolved cross-correlation approach for this study, because in spite of longer computational times it is more robust in the presence of non-Kolmogorov turbulence. However, we found that the model-based approach still provides useful information in detecting the existence of dome seeing or non-Kolmogorov turbulence which can be beneficial from an operational point of view.

### 3.2. Description of the $C_n^2$ and wind profiling method

The time-delayed cross correlation between two WFSs, $WFS_A$ and $WFS_B$, is described by the formula (Wilson, 2002; Wang *et al*, 2008; Cortes *et al*, 2012):

$$T^{AB}(\Delta u, \Delta v, \Delta t) = \frac{\left\langle \sum_{u,v} S_{u,v}^{A}(t) \cdot S_{u+\Delta u, v+\Delta v}^{B}(t+\Delta t) \right\rangle}{O^{AB}(\Delta u, \Delta v)} , \quad (2)$$

where $S_{u,v}^{WFS}(t)$ contains the X and Y slopes of the WFS in subaperture $(u,v)$ at time $t$, $\Delta u$ and $\Delta v$ are relative subaperture displacements in the WFS grid. The time delay of the measurement, $\Delta t$, is a multiple of the integration time that in our case ranges from 1/800 s to 0.4 s. $\sum_{u,v}$ denotes summation over all valid overlapping subapertures and $\langle \, \rangle$ represents the average over the time series.

Central to the temporal cross-correlation method is the $O^{AB}(\Delta u, \Delta v)$ matrix that contains the number of overlapping illuminated subapertures for an offset given by $\Delta u$ and $\Delta v$. It compensates for the effect of having fewer subapertures to correlate as one moves away from the centre of the correlation matrix (reduction in overlapping area). Figure 5 shows an example of this matrix, computed as the cross-correlation between the masks of WFS1 and WFS2 shown in Fig. 3. Notice the asymmetry present in the horizontal axis, which is due to the different masks used for each WFS.

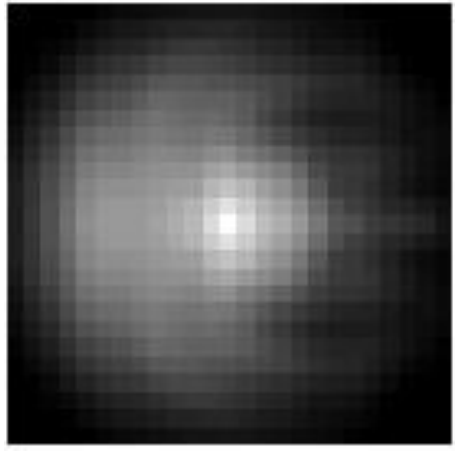

**Figure 5:** Compensation matrix $O^{12}$. An asymmetric pattern due to the different masking for invalid subapertures in each WFS is observed in the horizontal direction.

A two-dimensional deconvolution is applied (Wilson, 2002; Wang *et al*, 2008; Cortes *et al*, 2012) to the time delayed cross-correlation using the simultaneous autocorrelation of each WFS and the Fast Fourier Transform (FFT), i.e.

$$\mathrm{FT}^{-1}[\mathrm{FT}[T^{AB}]/\mathrm{FT}[A]] \quad , \quad (3)$$

where FT and FT$^{-1}$ are the Fourier Transform and inverse Fourier Transforms respectively, and $A$ is the average of the autocorrelations of WFS$_A$ and WFS$_B$:

$$A(\Delta u, \Delta v) = \frac{1}{2} \frac{\left\langle \sum_{u,v} S^A_{u,v}(t) \cdot S^A_{u+\Delta u, v+\Delta v}(t) \right\rangle}{O^{AA}(\Delta u, \Delta v)} + \frac{1}{2} \frac{\left\langle \sum_{u,v} S^B_{u,v}(t) \cdot S^B_{u+\Delta u, v+\Delta v}(t) \right\rangle}{O^{BB}(\Delta u, \Delta v)} \quad . \quad (4)$$

When $\Delta t$ in equation (2) is zero ($T^{AB}(\Delta u, \Delta v, 0)$) the peaks along the baseline connecting the two stars represent the turbulence in the corresponding bins. This provides an estimation of the altitude and strength of the turbulent layers, as illustrated in Fig. 6.

If the frozen flow hypothesis holds, (Taylor, 1938), the peaks in the correlation maps move with increasing $\Delta t$ depending on the wind and direction in each turbulent layer. A clear example of the latter is presented in the sequence of correlation images in Fig. 7.

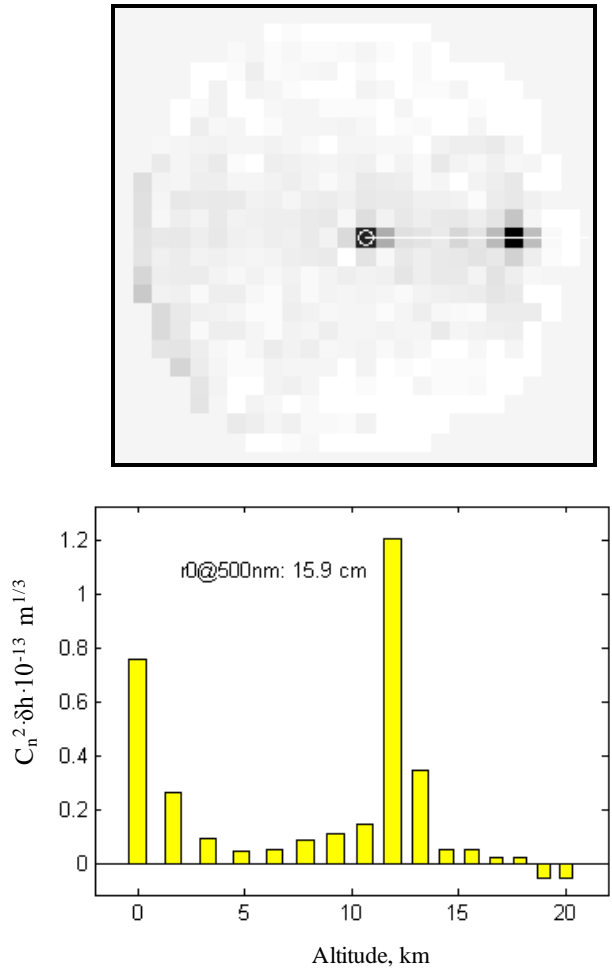

**Figure 6.** Profile resulting from the correlation between WFS1 and WFS2 averaged with the correlation between WFS4 and WFS3. The profile can be estimated by scanning the correlation matrix (top panel) along the baseline between the WFSs (horizontal axis in that case) for $\Delta t$ =0 s. The bottom bar plot shows the resulting profile (data from April 16$^{th}$, 2013).

By tracking the correlation peaks we can estimate not only the direction and speed of the wind, but also the $C_n^2$ value of the associated layer (Cortes *et al*, 2012).

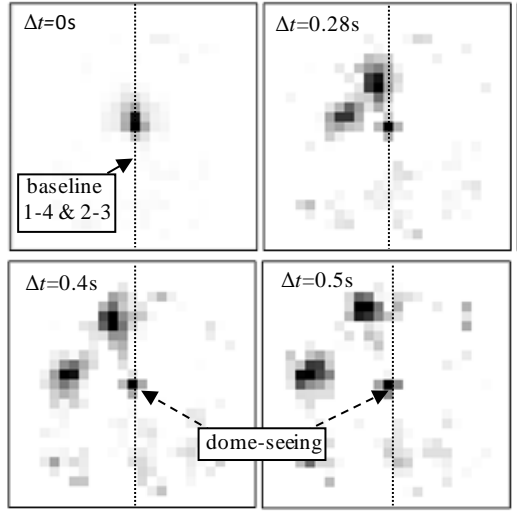

**Figure 7.** The sequence shows the time correlation from $\Delta t = 0$ to $\Delta t = 0.5$s, with two layers moving in different directions and a third static peak at the centre, corresponding to the dome seeing. The baseline corresponds to the vertical combinations of WFSs, i.e. 1-4 and 2-3 (on-sky data, April 15th, 2011).

## 4. DETERMINING THE DYNAMIC EVOLUTION OF THE TURBULENCE IN LAYERS

In this section we analyse the evolution of the turbulence in terms of its deformation in time for different altitudes and wind kinematics. Using the temporal correlation sequences as the one shown in Fig. 7, the layers can be tracked and their speed and direction estimated. As the individual correlation peaks are being tracked, the degree of correlation can also be computed, providing an estimation of the evolution of the layer, i.e. the correlation decay or the rate of "boiling" as it crosses the telescope's field of view.

Isolating the correlation peaks for tracking purposes is not trivial and it only works in situations where they can be individualized, i.e. they have different velocity vectors and are not overlapped along the tracks. A possibility to overcome this problem could be to use a detection algorithm as the one developed by Poyneer *et al* (2009) and used by Cortés *et al* (2013). It is out of the scope of our study to present a method for systematic identification of all wind-driven layers. Instead we try to characterise this evolution from carefully selected cases where the cross-correlation peaks are easily traceable.

By selecting cases where the turbulence layers can be clearly isolated and tracked we focus on the analysis and interpretation of the observed dynamics of the turbulence intensity. In particular, we have estimated the rate of de-correlation of the layered turbulence as it passes over the telescope.

### 4.1. Frozen flow assumption and the correlation decay rate

The spatio-temporal cross-correlation maps bring information upon the wind speed and direction of the frozen-flow layers, thanks to the correlation peaks moving across the maps. The evolution of the intensity and size of the cross-correlation peaks are of importance since the peak maximum intensity reveals how much of the turbulence has been perfectly translated by frozen flow at a given speed. Also, the broadening of the peak along time typically illustrates the wind shear phenomenon, accounting for the fact that we are not able to distinguish multiple layers of similar wind speeds inside a single altitude bin (Conan *et al*, 1995).

The latter broadening effect is exemplified in the next figure. We compare the evolution of an ideal layer simulated with the same altitude and velocity characteristics of a real one. A single layer at 4km is translated across the telescope pupil with the same speed and direction as the one determined from on-sky data. Pseudo-open loop slopes are produced for both cases, and they are used as an input for the wind-profiler data. Figure 8 shows the sequences of correlation images for the real (left column) and simulated (right) layers. Notice the broadening of the correlation peaks for the real case (left column) which does not occur in the ideal case.

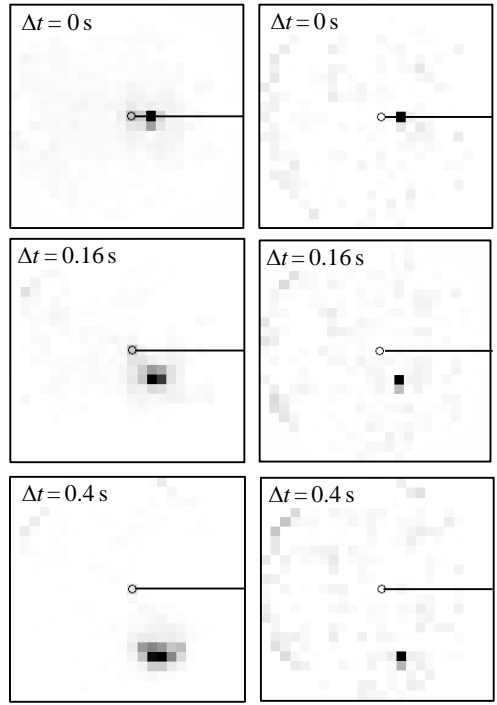

**Figure 8:** Sequence of cross-correlations between WFS1 and WFS2 for real data (November 7$^{th}$, 2012) and simulated turbulence. The wind speed and direction of the real turbulence (left sequence) have been used to generate the simulated case (right sequence).

The characterization of the exactly frozen flow (intensity evolution of the peak) could be obtained by tracking the point of maximum correlation. For instance, a two-dimensional Gaussian function can be fit to the correlation peaks for a later tracking of the Gaussian maximum along the temporal sequence. The free parameters for the Gaussian fit are: X-Y position, full width half maximum (FWHM), flux, maximum, ellipticity ratio and angle. This has been done for the same data in Fig. 8 and the results are presented in Fig. 9 (dotted line). The correlation values are normalized to the peak of the first correlation result in the sequence (peak at $\Delta t = 0$s).

The resulting curve of peak intensity decay can be highly noisy at the beginning of the sequence. This is due to the fact that the peak is initially narrow and the Gaussian is thus poorly sampled (the Gaussian fit gives a FWHM between 1 and 2 pixels at the start of the sequence.

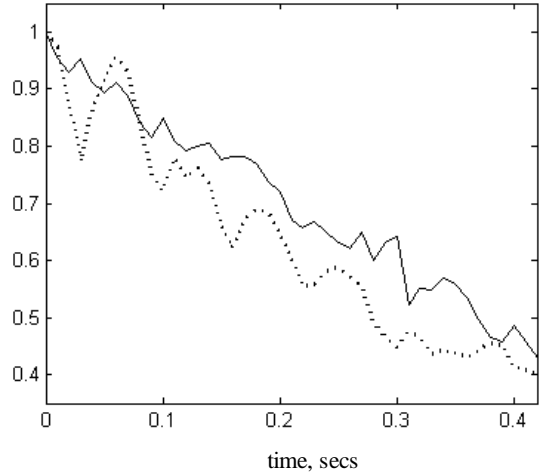

Figure 9. Example of two methods for estimating the decay ratio (continuous: 3x3 window; dotted: Gaussian fit)

The decay rate (slope of the curve) is more reliably obtained beyond $\Delta t = 0.1$ s. As a first conclusion, tracking the decay of the frozen flow using Gaussian fit on the correlation peaks is not easy. In practice, when the subaperture sampling is coarse (as in GeMS) it leads to difficult estimation of the decay rate. Secondly, this approach only accounts for the turbulence that moves with a single wind speed and direction.

Looking for an alternative to the Gaussian fit, we have tried using the average value of a 3x3 window around the maximum intensity. The decay of this integrated energy provides much smoother curves as observed in Fig. 9 (continuous line). In addition, the 3x3 pixel approach provides an estimate of the frozen flow decay rate, accounting for a certain dispersion of the wind speed and including it in the decay computation.

It can be argued that for large evolution times the energy around the peak maximum overflows the 3x3 window, but this situation is no longer representative of the initial layer wind speed. The choice of the averaging window size is arbitrary but at some point a decision has to be made on whether the more dispersive components are still part of a single layer. We think that, given the poor resolution of the cross-correlation maps, a 3x3 matrix allows to gather those turbulence components that belong to the same layer. This method is used in the reminder of the paper to estimate the decay ratio.

A final sanity check is performed on this method by comparing the real and the perfectly theoretical layers at 4Km shown in Fig. 8. If the method works correctly an approximately constant curve should be obtained for the ideal case (no decay). As expected, Fig. 10 (dashed line) shows no decay for this case as long as a fraction of the initial layer stays inside the pupil ($\Delta t < 0.5$s). For the on-sky data a clear constant decorrelation rate is observed (Fig. 10, continuous line).

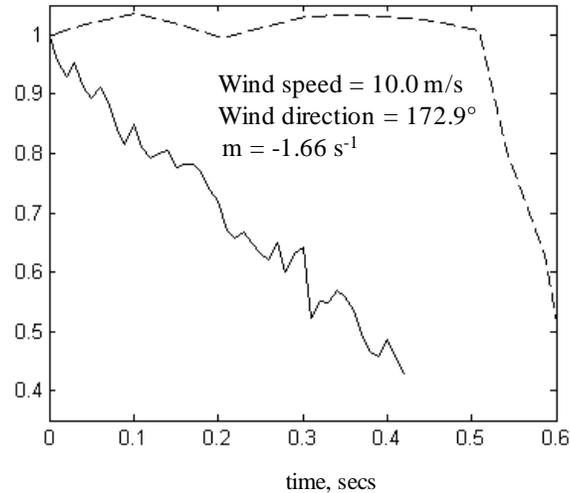

**Figure 10:** Decay ratios for simulated and real data. The simulated parameters are defined similar to those of the real turbulence (wind speed and direction). Dashed line: simulated turbulence; Continuous line: real data from November 7th, 2012.

This wind profiling method was extensively tested in several campaigns during 2011, 2012 and 2013. An example is presented in Fig. 11 where the profiles for a complete night (April 16th, 2013) are shown. A strong ground layer and another at 11 Km (jet stream) are clearly identifiable.

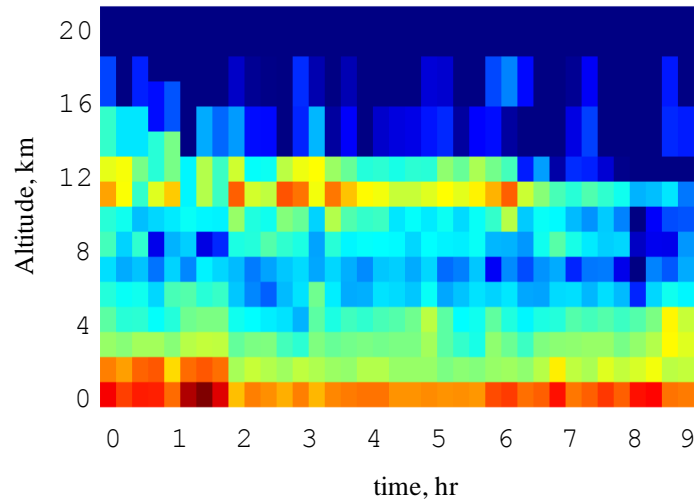

**Figure 11.** Turbulence profile (logarithmic scale) for a complete night (April 16th, 2013). The jet stream at around 22 Km is clearly visible.

The data collected during these campaigns, allowed us to gather 74 sequences of turbulence layers where their decay rate could be determined. The set includes different conditions of altitudes, strength and velocities. In the case of the wind speed, their values ranged from 0 m/s for strong dome seeing conditions (9 cases) to speeds up to 24 m/s. Higher speeds were detected above 12 Km, but their translation time across the telescope's field of view was not sufficient as to reliably estimate their velocity.

Next, we separate the analysis in three distinctive cases according to the altitudes of the layers of interest: i) turbulence inside the dome; ii) turbulence on the ground but above the telescope (with clear translation speeds); iii) middle and high altitude layers (between 3 and 20 Km) including those related to the jet stream.

### 4.2. Frozen flow for turbulence inside the dome

As mentioned before, turbulences inside the dome are fairly common during operation of large telescopes. We have found that this type of turbulence tends to remain stationary with zero translation velocity. The dome turbulence is spatially located in the same place in all WFSs, so it is seen as a layer at the ground. Hence, the profiler assigns its energy to the first bin and as this phenomenon is quasi static a distinctive stationary peak will appear during the temporal correlation sequence.

Fig. 12 (top) shows the profile for a case with strong dome turbulence and in the bottom panel, the intensity of the central pixel in the sequence of time delayed correlation is shown. The plot (normalized to its value at $\Delta t = 0$) shows a rapid initial decay, that gradually stabilizes to a rate of $-0.32 s^{-1}$. This fast initial decay was found to be caused by a ground layer passing over the telescope that at the start of the temporal correlation is merged with the dome seeing. As time passes, this part of the turbulence moves away from the central peak so its correlation with the fixed frame is lost, but the dome turbulence will evolve much slower, maintaining a positive match for a longer time. By measuring the slope of the asymptotic dashed line, the rate of de-correlation can be estimated.

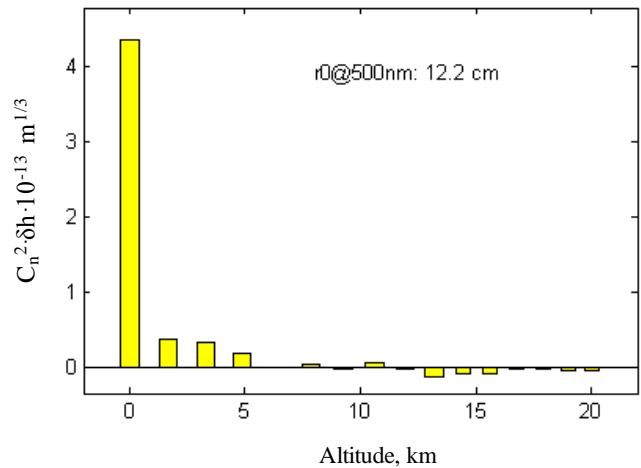

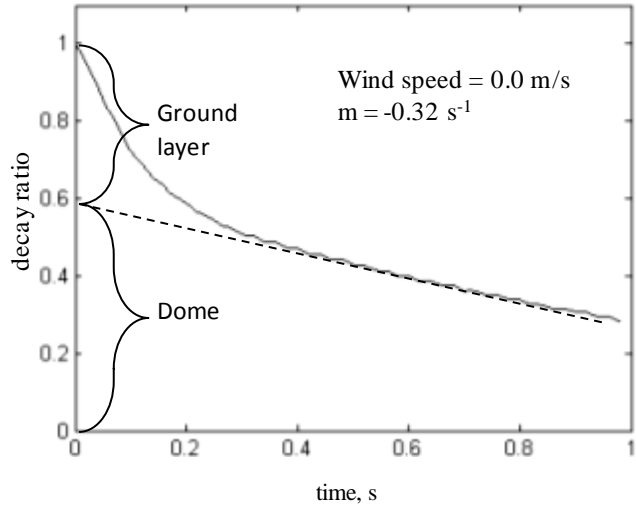

**Figure 12:** Top: estimated turbulence profile in altitude, showing a strong component at the ground layer; Bottom: decorrelation rate of turbulence inside the dome (dashed line) and separation from the ground layer turbulence above the telescope (data from November 7$^{th}$, 2012)

This procedure offers a powerful tool during operation of the telescope by estimating the dome seeing at any time, triggering the corrective measures to counteract this problem.

### 4.3. Frozen flow for turbulence at the ground layer

Similar to the previous case, Fig. 13 (top) shows a profile with most of the turbulence at the ground layer. Here, the sequence of correlations show that the main layer moves at a speed of 8.8 m/s and is located at an altitude between 0 and 1.7 Km. By tracking the correlation peak and measuring its energy, a relative decay of -1.33 s$^{-1}$ is observed before the correlation disappears.

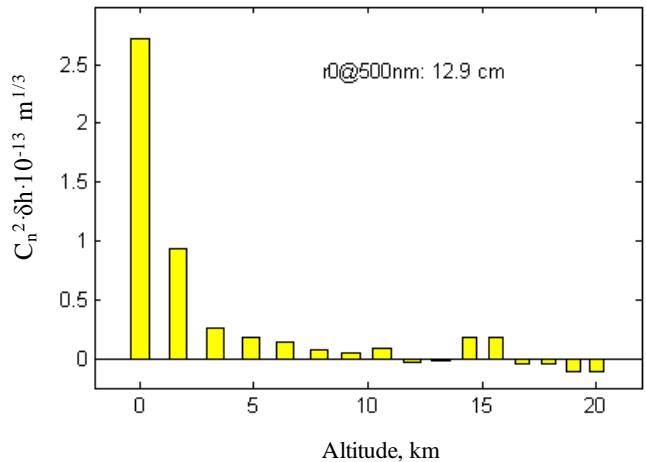

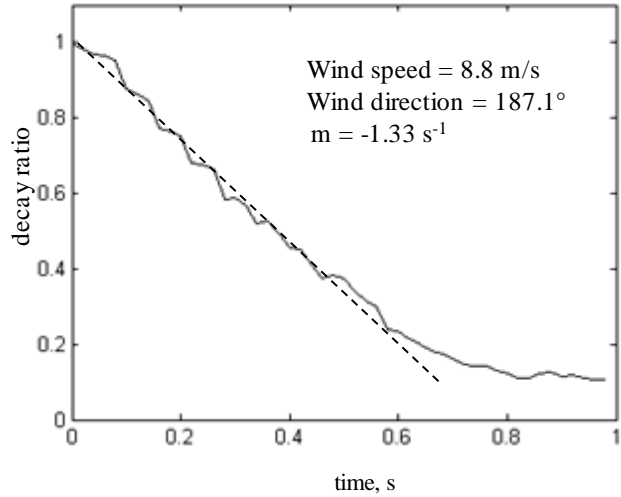

**Figure 13:** Top: ground layer turbulence with absence of dome seeing; Bottom: estimation of the turbulence decay rate (dashed line). Total absence of dome seeing (data from June 12th, 2013)

### 4.4. Frozen flow for turbulence at high altitudes

For higher altitude cases, we started with the profile in Fig. 14 showing a strong turbulence in the third bin (between 3.3 and 4.9 Km). Tracking the strongest correlation peak for this case, a speed of 10.0 m/s is measured with a decay rate of -1.66 $s^{-1}$, as already shown in Fig. 9 and Fig. 10.

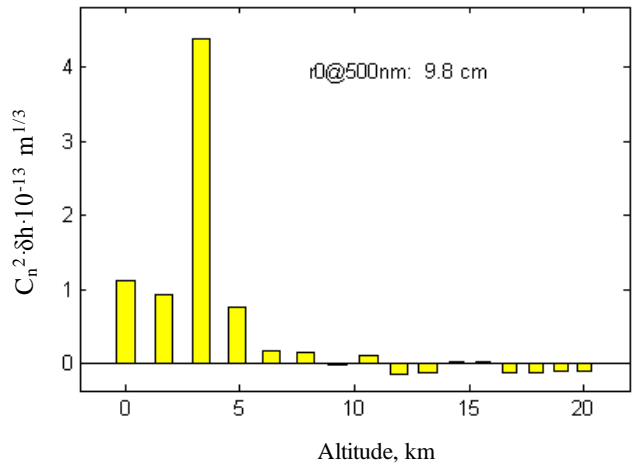

**Figure 14:** Profile for a turbulence concentrated in the third bin corresponding to altitudes between 3.3 and 4.9 Km (data from November 7th, 2012)

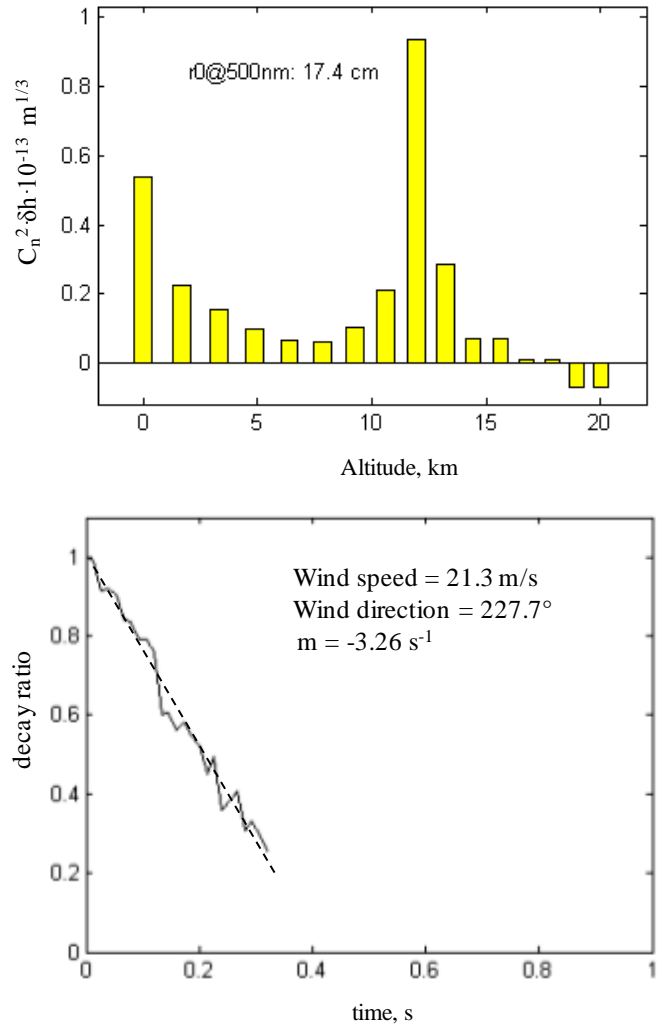

**Figure 15:** Top: profile for a turbulence concentrated in the ninth bin (11.9 – 13.2 Km) from data obtained during data from April 17$^{th}$, 2013; Bottom: estimation of the turbulence decay rate (dashed line)

Finally, the profile in Fig. 15 (top panel) presents a typical case of a strong layer corresponding to the jet stream at an altitude of around 12 Km. The measured speed is 21.3 m/s with a decay ratio of -3.26 s$^{-1}$. A steeper decay in the cross-correlation is observed in this case (Fig. 15, bottom panel) when compared to the previous cases.

### 4.5. Dependence of frozen flow to wind speed

The most interesting result of this turbulence "boiling" comes from an integrated analysis for several cases collected at Gemini South during the last two years. Fig. 16 plots the relative decay ratios against wind speed under different turbulence conditions and time.

A clear pattern can be observed between the dissipation rate and the wind speed with a remarkable resemblance to a linear dependence. Fitting a linear function to the set of points gives a slope of about -6.4 m-1. This suggests that the 'melting' of the turbulence for short time scales, depends on the distance travelled by the layer rather than its speed, strength or altitude. Hence, the turbulence not only translates but also deforms and the faster the layer travels, the faster it decorrelates. This result could be used to build models of the turbulence evolution including both frozen flow and melting (e.g. Berdja A. and Borgnino J. 2007), however so far, only few data analysis are available considering frozen flow and melting together. A somewhat similar analysis has been carried out in Schöck and Spillar (2000). They also found similar trends for the decorrelation with time, and show that the frozen flow hypothesis holds for period of tens of milliseconds. They do not seem to find a linear dependence with speed, however, these earlier results were obtained using a single WFS experiment, and only for a few data sets. Further study with more data will be required to confirm and explain the linear behaviour that we evidence with GeMS data.

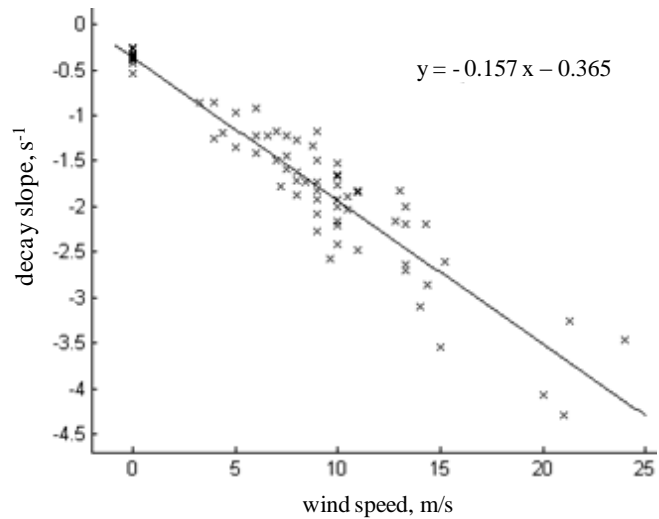

**Figure 16:** Rate of decay in the cross-correlation of slopes. The regression shows a clear linear dependence between wind speed and the dissipation of turbulence correlation. The slope of the linear fit is in units of $m^{-1}$, suggesting that this de-correlation can be expressed in terms of the distance travelled by the layers across the field of view.

Fig. 16 also highlights the fact that many layers have been found with decay rate between -1 $s^{-1}$ and -3 $s^{-1}$. Considering for instance a layer with a decay rate of -2 $s^{-1}$, which occurs for a wind speed around 10m/s, the correlation of this layer with the shifted one that occurred 250ms earlier (more than 125 AO cycles) still contains half of the energy in the original correlation peak. For all the cases presented in Fig. 16 the decay rate is smaller than -5 $s^{-1}$, meaning that the correlation decreases by a factor lower than 2 after 100ms, i.e. more than 50 cycles of the GeMS AO loop.

Another way to emphasize the importance of this result is to mention that after the usual 2-frame delay of the AO loop, i.e. the time the prediction is mostly required (around 5ms), the correlation of an individual layer under frozen-flow is always maintained to levels higher than 97.5%. And for the layers with the most commonly detected wind speed in Fig. 16, i.e. around 10 m/s, the intensity of the correlation peak is maintained to 99.5% of its original intensity.

The relationship between the decay rate and the wind speed can now be used to construct a predictive controller. The temporal evolution of the turbulent phase in the identified layers can now be modelled by the phase continuous function in space and time $\varphi$, according to

$$\varphi(x,y,t) = \left(1 + r(|v|)(t-t_0)\right) \cdot \varphi\left(x - v_x(t-t_0), y - v_y(t-t_0), t_0\right) + \varepsilon(x,y,t) \quad (5)$$

where $r$ is the decay rate and instants $t$ and $t_0$ are assumed to be close enough so that the reduction product $r(|v|)(t-t_0)$ has a magnitude lower than unity. In addition, $\varepsilon$ is a coloured noise maintaining the statistical properties of $\varphi$ (Kolmogorov or von Kármán for instance), representing the effect of the wind-shear or buoyancy of the turbulence. To build such temporal model, the decay rate could either be taken from the linear fit in Fig. 16 as a function of the wind speed $|v|$ or if the decay analysis of the considered layer is available as in Figures 9, 12, 13 and 15, from measurements of the intensity decrease. Models similar to equation (5) have already been used in the literature to design adaptive optics controllers with Kalman filter and a first-order auto-regressive model (Poyneer *et al*, 2009; Gilles *et al*, 2013). The value of this factor has been assumed 1.0 (Gilles *et al*, 2013) or 0.995 and 0.99 (Poyneer *et al*, 2009).

The analysis presented here highlights that the factor should rather be a function of the decay rate and the delay, although quantitatively the values obtained may stay very close to 1 or 0.995.

## 5. CONCLUSIONS

We have used the GeMS turbulence profiler to characterize the evolution of the turbulence layers. This tool relies on the well-known SLODAR technique but modified to deal with laser-based data with fratricide effect and closed-loop AO measurements. The data gathered over a two-year period have shown the frequent presence of a strong dome turbulence component. We identified that this turbulence does not follow Kolmogorov or Von Karman statistics. This led us to adopt a deconvolution approach to analyse the spatio-temporal slopes correlations.

The principal results from this analysis refer to the subject of the frozen-flow assumption. From our wind profile analysis, we show how the decay rate of the frozen-flow correlation of each layer can be estimated and more important, we have found a linear relationship between the frozen-flow decay rate and the translational speed of the layers. This slow de-correlation also confirms the potential of predictive control techniques in wide-field AO systems, as these results can be included in the temporal modelling of the phase evolution.

Some improvements to this turbulence characterization tools should be addressed in further work. In particular, the development of a more automatic way to track the correlation peaks, especially when multiple peaks coincide along the temporal sequence can be designed. A combination of the method described above with a recent detection method applied to a Fourier analysis (Poyneer *et al*, 2009; Ammons *et al*, 2013; Cortes *et al*, 2013) may provide a solution.

## ACKNOWLEDGMENTS

This work was supported by the Chilean Research Council (CONICYT), grant Fondecyt 1120626, and by the French ANR program WASABI-ANR-13-PDOC-0006-01.

## REFERENCES

Ammons S. M., Poyneer L., Gavel D., Macintosh B., Kupke R., Max C., Rockosi C., Srikar S., Rudy A., Neichel B., 2013, in Esposito S., ed., Proc. AO4ELT3 Conf., available online at: http://ao4elt3.sciencesconf.org/
Berdja A. and Borgnino J., 2007, MNRAS, 378, 1177
Britton M. C., 2004, Proc. SPIE 5497, 290
Butterley T., Wilson R. W. and Sarazin M., 2006, MNRAS, 369, 835
Conan J.M., Rousset G. and Madec P.Y., 1995, J. Opt. Soc. Am. A, 12, 1559
Cortés A., Poyneer L., Neichel B., Mark Ammons M., Rudy A., Guesalaga A., 2013, in Esposito S., ed., Proc. AO4ELT3 Conf., available online at: http://ao4elt3.sciencesconf.org/
Cortés A., Neichel B., Guesalaga A., Osborn J., Rigaut F. and Guzmán D., 2012, MNRAS, 427, 2089
Fusco T. and Costille A., 2010, Proc. SPIE 7736, 77360J
Gavel D. T. and Wiberg D., 2002, Proc. SPIE 4839, 890
Gilles L. and Ellerbroek, B., 2010, J. Opt. Soc. Am. A, 27, 76
Gilles L., Massioni P., Kulcsar C., Raynaud H.-F. and Ellerbroek B., 2013, J. Opt. Soc. Am. A, 30, 898
Guesalaga A., Neichel B., Cortés A., Guzmán D., 2013, in Esposito S., ed., Proc. AO4ELT3 Conf., available online at: http://ao4elt3.sciencesconf.org/
Hinnen K., Verhagen M., and Doelman N., 2010, J. Opt. Soc. Am. A, 24, 1714
Jolissaint, L., 2006, PASP, 118, 1205
Le Roux B., Conan J.-M., Kulcsar C., Raynaud H.-F., Mugnier L. M., and Fusco T., 2004, J. Opt. Soc. Am. A 21, 1261
Neichel B. et al., 2010, in Ellerbroek B. L., Hart M., Hubin M., Wizinowich P.L., eds. Proc. SPIE Conf., 7736, 773606
Neichel B. et al, 2011, in Veran J. P., ed., Proc. AO4ELT2 Conf., available online at: http://ao4elt2.lesia.obspm.fr/
Neichel B., et al., 2012, in , Ellerbroek B.L., Marchetti E., Véran J.-P., eds. Proc. SPIE Conf., 8447, 84474Q
Osborn J., De Cos Juez F. J., Guzman D., Butterley T., Myers R., Guesalaga A, and Laine J., 2012, Opt. Express, 20, 2420
Poyneer L., Macintosh B. A., and Véran J.-P., 2007, J. Opt. Soc. Am. A , 24, 2645
Poyneer L., Van Dam M. and Véran J. P., 2009, J. Opt. Soc. Am. A, 26, 833
Rigaut F., Neichel B., Boccas M., *et al.*, 2014, MNRAS, 437, 2361
Saint-Jacques, D. and Baldwin, J. E., 2000, in Léna, P. and Quirrenbach A. eds., Proc. SPIE Conf., 8006, 951
Schöck M. and Spillar E. J., 2000, J. Opt. Soc. Am. A, 17, 1650
Taylor G. I., 1938, Proc. R. Soc., Ser. A, 164, 476
Wang L., Schöck M. and Chanan G., 2008, Appl. Opt., 47, 1880
Wilson R. W., 2002, MNRAS, 337, 103